\documentclass[twocolumn]{aastex62}

\usepackage{color}

\usepackage{amsmath}
\usepackage{comment}

\newcommand{\HI}{H{\sc~i}}
\newcommand{\HII}{H{\sc~ii}}
\newcommand{\HeI}{He{\sc~i}}
\newcommand{\HeII}{He{\sc~ii}}
\newcommand{\HeIII}{He{\sc~iii}}

\graphicspath{{./}{figures/}}

\received{XXX}
\revised{YYY}
\accepted{ZZZ}

\begin{document}

\title{Non-equilibrium temperature evolution of ionization fronts during the Epoch of Reionization}

\author[0000-0002-0786-7307]{Chenxiao Zeng}
\affiliation{Center for Cosmology and Astroparticle Physics (CCAPP)\\
The Ohio State University, Columbus, OH 43210, USA}
\affiliation{Department of Physics,  The Ohio State University, Columbus, OH 43210, USA}

\author[0000-0002-2951-4932]{Christopher M. Hirata}
\affiliation{Center for Cosmology and Astroparticle Physics (CCAPP)\\
The Ohio State University, Columbus, OH 43210, USA}
\affiliation{Department of Physics, The Ohio State University, Columbus, OH 43210, USA}
\affiliation{Department of Astronomy, The Ohio State University, Columbus, OH 43210, USA}

\begin{abstract}

The epoch of reionization (EoR) marks the end of the Cosmic Dawn and the beginning of large-scale structure formation in the universe. The impulsive ionization fronts (I-fronts) heat and ionize the gas within the reionization bubbles in the intergalactic medium (IGM). The temperature during this process is a key yet uncertain ingredient in current models. Typically, reionization simulations assume that all baryonic species are in instantaneous thermal equilibrium with each other during the passage of an I-front. Here we present a new model of the temperature evolution for the ionization front by studying non-equilibrium effects. In particular, we include the energy transfer between major baryon species ($e^{-}$, \HI, \HII, \HeI, and \HeII) and investigate their impacts on the post-ionization front temperature $T_{\mathrm{re}}$. For a better step-size control when solving the stiff equations, we implement an implicit method and construct an energy transfer rate matrix. We find that the assumption of equilibration is valid for a low-speed ionization front ($\lessapprox\ 10^9~\mathrm{cm}/\mathrm{s}$), but deviations from equilibrium occur for faster fronts. The post-front temperature $T_{\mathrm{re}}$ is lower by up to 19.7\% (at $3\times 10^9$ cm/s) or 30.8\% (at $10^{10}$ cm/s) relative to the equilibrium case.

\end{abstract}


\section{Introduction} \label{sec:intro}
The Epoch of Reionization (EoR) is the process when ultraviolet (UV) photons from the first stars ionized almost all of the neutral hydrogen in the intergalactic medium (IGM). This transition is thought to occur around $z \sim 6 - 12$ \citep{2001AJ....122.2850B, 2006AJ....132..117F, 2015MNRAS.447..499M, 2018arXiv180706209P}.

Temperature evolution of the IGM is one of the key ingredients in any reionization model, constraining the source energy and injection processes into the IGM. This includes both the standard astrophysical sources responsible for hydrogen and helium reionization, as well as potential new physics such as dark matter-baryon interactions \citep{2017JCAP...11..043M}. Understanding the IGM temperature is also a key element in cosmological constraints with the Lyman-$\alpha$ forest \citep[e.g.][]{2004MNRAS.354..684V, 2005ApJ...635..761M, 2006MNRAS.365..231V, 2015JCAP...11..011P, 2019JCAP...07..017C}, and the thermal evolution shortly after reionization is particularly important for constraints on warm dark matter \citep[e.g.][]{2008PhRvL.100d1304V, 2013PhRvD..88d3502V, 2016JCAP...08..012B, 2017JCAP...12..013B, 2017PhRvD..96b3522I, 2017PhRvL.119c1302I, 2017MNRAS.471.4606A, 2019arXiv191209397G}. In the standard theory, as ionization fronts (I-fronts) propagate outward from ionization sources, a sharp boundary between ionized particles and neutrals forms \citep{2016ARA&A..54..313M}. As I-fronts passing through, the IGM gas are rapidly heated to the order of $10^4$K \citep{1994MNRAS...266...343M, 1997MNRAS.292...27H, 2018MNRAS...474...2173H, 2019ApJ...874..154D} which is so-called post-ionization-front temperature $T_{\mathrm{re}}$. The heated IGM then undergoes a cooling process by the joint effects of adiabatic expansion of the universe and inverse Compton scattering with Cosmic Microwave Background (CMB) photons. It also undergoes a complicated hydrodynamic relaxation process as pre-existing small-scale structures in the IGM are disrupted by the increase in temperature \citep[e.g.][]{2004MNRAS.348..753S, 2005MNRAS.361..405I, 2018MNRAS...474...2173H, 2020arXiv200202467D}.

There is a great uncertainty on the post-ionization temperature $T_{\mathrm{re}}$, due to the difficulty of simulating and measuring the thermal evolution history. Experiments estimate the volume-weighted mean temperature of the IGM during EoR by measuring thermal broadening of Lyman-alpha (Ly$\alpha$) forest absorption features \citep{2000MNRAS.318..817S, 2010ApJ...718..199L, 2011MNRAS.410.1096B, 2012MNRAS.424.1723G, 2012ApJ...757L..30R, 2014MNRAS.441.1916B, 2014MNRAS.438.2499B}. An example of simulating $T_{\mathrm{re}}$ is shown in \citet{2019ApJ...874..154D}, where they used high-resolution radiative transfer (RT) simulations. In their work, $T_{\mathrm{re}}$ mildly depends on incident spectrum and primarily sensitive to the I-front speeds. One approximation they made is that all baryon species reach equilibration states instantaneously as I-front passing through. This is assuming the timescale of thermalizing baryon species other than photoelectrons is small compared with the time that gas stays in the I-front. However, it is possible that non-equilibrium effects can influence the thermal evolution in the I-front. This is because if the front speed is large enough, the local gas density is low and the energy-transfer interaction rates become comparable with the timescale that gas stay in the I-front. \citet[Appendix B]{2019ApJ...874..154D} provided order-of-magnitude estimates arguing that the equilibrium approximation should hold for most of the parameter space of I-front speeds during reionization.

In this paper, we build a non-equilibrium model of $T_{\mathrm{re}}$ by coupling an implicit stiff solver to the 1-dimensional grid-based I-front model of \citet{2018MNRAS...474...2173H}. We solve a set of stiff equations describing the energy transfer between species after photoelectrons are thermalized at each time step. Allowing for non-equilibrium effects, both $T_{\mathrm{re}}$ and equilibrium states in the I-front are affected. We find that at the midpoint of the ionization front (where $x_e\sim 0.5$) the temperatures of the species are in equilibrium in most of the parameter space. However in the mostly neutral region, there can be significant deviations from equilibrium, and this affects the cooling rate and final temperature if the ionization front is very fast. The source code is placed in a public GitHub repository\footnote{https://github.com/frankelzeng/reionization}.

This paper is structured as follows. In Section \ref{sec:model}, we briefly review the one-dimensional grid model. In Section \ref{sec:stiff}, we present our stiff solver algorithm in detail and explain the energy transfer cross sections between species in four categories. In Section \ref{sec:result}, we build the dependence of $T_{\mathrm{re}}$ on the incident blackbody spectrum $T_{\mathrm{bb}}$ and I-front speed $v_i$ when non-equilibrium effects are present. We conclude in Section \ref{sec:conc}.

\section{Method: Grid Model} \label{sec:model}

Here we review the grid model for temperature evolution during the epoch of reionization. We only summarize the main results of the model, and detailed derivations are explained in the appendix of \citet{2018MNRAS...474...2173H}. Following the physical reasoning of \citet{1994MNRAS...266...343M}, this model describes density-dependent reionization temperature $T_{\mathrm{re}}$.

The model is a time-dependent ionization front in one dimension with the depth parameter $N_\mathrm{H}$ (units: $\mathrm{cm}^{-2}$) the total hydrogen column. The ionization front is built on a grid of $N_{\mathrm{grid}}$ cells of width $\Delta N_\mathrm{H}$, and each cell $j\in \{0,...,N_{\mathrm{grid}}\}$ contains a hydrogen neutral fraction $y_{\mathrm{HI},j}$, a helium neutral fraction $y_{\mathrm{HeI},j}$, and an energy per hydrogen nucleus $E_j$. We consider only the first ionization front, i.e., \HI/\HeI$\rightarrow$\HII/\HeII; the \HeII$\rightarrow$\HeIII\ ionization occurs later (see, e.g., the discussion in the review by \citealt{2016ARA&A..54..313M}) and is not treated in this paper.

A flux of photons $F$ (units: photons cm$^{-2}$ s$^{-1}$) is incident on the left side of the grid. The ionization front has velocity $v_i = F(1 - v_i/c)/[n_\mathrm{H}(1 + f_\mathrm{He})]$, where $n_\mathrm{H}$ is the three-dimensional hydrogen number density and $f_\mathrm{He}$ is the helium-to-hydrogen ratio. The relativistic correction term $v_i/c$ accounts for the finite time that incident photons travel to the I-front. To visualize the temperature in terms of the total hydrogen column $N_\mathrm{H}$, we introduce a rescaled time $t'\equiv Ft$ (units: photons cm$^{-2}$).

Breaking the incident flux into a set of wavelength bins $\lambda$ with each bin contains a fraction $f_\lambda$ of the photon flux, the photoionization rates for hydrogen and helium are
\begin{eqnarray}
    \frac{d y_{\mathrm{HI}, j}}{d t^{\prime}} & = & \sum_{\lambda} A_{j \lambda} \frac{\tau_{j \lambda}^{\mathrm{HI}}}{\tau_{j \lambda}^{\mathrm{HI}}+\tau_{j \lambda}^{\mathrm{HeI}}}\label{eqn:yHI} ~~~{\rm and}\nonumber \\
    \frac{d y_{\mathrm{HeI}, j}}{d t^{\prime}} & = & \sum_{\lambda}\frac{1}{f_\mathrm{He}} A_{j \lambda} \frac{\tau_{j \lambda}^{\mathrm{HeI}}}{\tau_{j \lambda}^{\mathrm{HI}}+\tau_{j \lambda}^{\mathrm{Hel}}}\label{eqn:yHeI} ,
\end{eqnarray}
where $\tau_{j \lambda}^{\mathrm{HI}} = \Delta N_\mathrm{H} y_{\mathrm{HI},j} \sigma_\alpha^\mathrm{HI}/(1 - v_i/c)$ and $\tau_{j \lambda}^{\mathrm{HeI}} = f_\mathrm{He} \Delta N_\mathrm{H} y_{\mathrm{HeI},j} \sigma_\alpha^\mathrm{HeI}/(1 - v_i/c)$ are optical depths to photons in frequency bin $\lambda$. The same relativistic correction term as in $F$ indicates that the grid width should be broaden for I-front moving close to the speed of light. Here ${\cal A}_{j\lambda}$ is the number of absorbed photons per hydrogen nucleus per rescaled time, given by
\begin{equation}
    {\cal A}_{j\lambda} = f_{\lambda} \exp \left(-\sum_{j^{\prime}=0}^{j-1} \tau_{j^{\prime} \alpha}\right) \frac{1-\exp \left(-\tau_{j \alpha}\right)}{\Delta N_{\mathrm{H}}}.
\end{equation}

The photoionization and collisional cooling lead to a net heating rate
\begin{equation}
\label{eqn:Ej}
    \begin{split}
        \frac{dE_j}{dt'} = & \sum_{\lambda} \mathcal{A}_{j \lambda} \frac{\tau_{j \lambda}^{\mathrm{HI}}\left(h \nu_\lambda -I_{\mathrm{HI}}\right)+\tau_{j \lambda}^{\mathrm{HeI}}\left(h \nu_\lambda -I_{\mathrm{HeI}}\right)}{\tau_{j \lambda}^{\mathrm{HI}}+\tau_{j \lambda}^{\mathrm{Hel}}} \\
        & -\frac{y_{\mathrm{HI}, j} x_{e, j} (1 - v_i/c)}{v_{i}\left(1+f_{\mathrm{He}}\right)} \sum_{n=2}^{3} q_{1 \rightarrow n} h \nu_{1}\left(1-n^{-2}\right),
    \end{split}
\end{equation}
where $E_j$ is the thermal energy per hydrogen nucleus in grid cell $j$; $x_{e,j}$ is the energy-temperature conversion factor defined in Table \ref{tab:f_alpha}; $\nu_\lambda \equiv c/\lambda$ is the frequency at each bin; $q_{1 \rightarrow n}$ represents the collisional excitation rate coefficient (units: cm$^3$s$^{-1}$; we use coefficients from \citealt{1983MNRAS.202P..15A}) for exciting a hydrogen atom to the $n^{\rm th}$ level; and $I_\mathrm{HI}$ and $I_\mathrm{HeI}$ are the ionization energies.

The major baryonic components of the IGM are $e^{-}$, \HI\ , \HII\ , \HeI\ , and \HeII. In what follows, we use the subscripts $\alpha,\beta,...$ to denote these components. We write the energy in component $\alpha$ per hydrogen nucleus as $E_\alpha$, and its temperature as
\begin{equation}
T_\alpha = \frac{2E_\alpha}{3k_{\rm B}f_\alpha},
\end{equation}
where $f_\alpha = n_\alpha/n_{\rm H}$ is the abundance of that species relative to total hydrogen nuclei. These factors are listed in Table \ref{tab:f_alpha}. The assumption of equilibrium states that all these temperatures are equal, in which case
\begin{equation}
T = \frac{2E}{3k_{\rm B} \sum_\alpha f_\alpha}.
\end{equation}

When a neutral atom is ionized, its kinetic energy becomes that of the ion, so we include a kinetic energy transfer from \HI\ to \HII\ and from \HeI\ to \HeII inversely proportional to the neutral fraction.

\begin{table}
\centering
\caption{Energy-temperature conversion factors.} \label{tab:f_alpha}
\begin{tabular}{ccccc}
\tablewidth{0pt}
\hline
\hline
$e^-$: $f_1$ & \HI: $f_2$ & \HII: $f_3$ & \HeI: $f_4$ & \HeII: $f_5$\\
\hline
$x_e$ & $y_{\mathrm{HI}}$     & $1-y_{\mathrm{HI}}$    & $f_{\mathrm{He}}y_{\mathrm{He}}$ & $f_{\mathrm{He}}(1 - y_{\mathrm{He}})$\\
\hline
\multicolumn{5}{c}{{\em Note} --- For electrons, $x_e = 1-y_{\mathrm{HI}} + f_{\mathrm{He}}(1 - y_{\mathrm{He}})$.}
\end{tabular}
\end{table}

\section{Method: Stiff Solver for Interactions} \label{sec:stiff}
We extend the equilibrium (electron-only) grid model by investigating the temperature evolution of five species ($e^{-}$, \HI\ , \HII\ , \HeI\ , and \HeII) separately. This requires us to study the mutual interactions and energy exchange within different species in the IGM. We only include two-body interactions, which can be represented by ${5\choose 2} = 10$ interaction rates. The system of equations is considered stiff because some of the energy transfer rates are large compared with the time scale of the overall reionization process, indicating an unacceptably small step size when doing integration. In this work, we construct an implicit stiff solver to reflect the temperatures evolution. We make an approximation that the species equilibrate among themselves immediately because this process is much faster than that within different components (e.g., electron-ion or ion-ion is faster than electron-ion).

Under non-equilibrium conditions, the components of the plasma have different temperatures. When time is rescaled to $t'$ and assuming there is no relative drift, the equilibration is described by \citep{Anders1990}
\begin{equation}
    \label{eqn:equi}
    \frac{dT_\alpha}{dt'} = \frac{dt}{dt'}\frac{dT_\alpha}{dt} = \frac{1 - v_i/c}{v_i n_\mathrm{H}(1+f_{\mathrm{He}})} \sum_{\beta\neq\alpha}\nu_{\alpha\beta}(T_\beta - T_\alpha),
\end{equation}
where $\nu_{\alpha\beta}$ is the energy transfer rate between component $\alpha$ and $\beta$ in physical time. Conservation of energy in the transfer requires the symmetry relation
\begin{equation}
f_\alpha \nu_{\alpha\beta} = f_\beta \nu_{\beta\alpha}.
\label{eq:sym}
\end{equation}
We express the temperature of each IGM component in an array $\mathbf{T}$ and the energy transferring rate in a matrix $\mathbf{M}$. The stiff equilibration can be converted to a matrix operation
\begin{equation}
\label{eqn:stiff}
\begin{split}
    \mathbf{T}(t' + \Delta t') & = \mathbf{T}(t') + \Delta t' \cdot \mathbf{M} \mathbf{T}(t' + \Delta t') \\
    & =  (\mathbf{I} - \Delta t' \cdot \mathbf{M})^{-1} \mathbf{T}(t'),
\end{split}
\end{equation}
where $\mathbf{I}$ is the $5\times5$ identity matrix, $\mathbf{T}$ has components $T_\alpha$, $\mathbf{M}$ has entries of some functions of the interaction rate $\Tilde{\nu}_{\alpha\beta}$, and $\alpha,\beta \in \{1,2,...,5\}$ denote each IGM species in the order of $e^{-}$, \HI, \HII, \HeI, and \HeII. We initialize the temperature of all species with $10^{-6}\ $K, which is ten orders lower than $T_{re}$, and the final result is insensitive to this initialization. The hydrogen and helium neutral fractions are both initialized to unity. In the code, we inserted hydrogen ($\sim 10^{-4}$) and helium residuals ($\sim 10^{-7}$) before reionization for numerical reasons; these do not affect the final temperature \citep{2011PhRvD..83d3513A}.

Using Equation~(\ref{eqn:stiff}), it is straightforward to show that $\mathbf{M}$ has the form
\begin{equation}
    \label{eqn:M}
    \mathbf{M} = \left(
\begin{array}{ccccc}
        \!\!\displaystyle{-\sum_{\beta\neq 1}} \nu_{1\beta}\!\!\! & \nu_{12} & \nu_{13} & \nu_{14} & \nu_{15}\\ 
        
        \nu_{21} & \!\!\!\displaystyle{-\sum_{\beta\neq 2}}\nu_{2\beta}\!\!\! & \nu_{23} & \nu_{24} & \nu_{25}\\ 
        
        \nu_{31} & \nu_{32} & \!\!\!\displaystyle{-\sum_{\beta\neq 3}}\nu_{3\beta}\!\!\! & \nu_{34} & \nu_{35}\\
        
        \nu_{41} & \nu_{42} & \nu_{43} & \!\!\!\displaystyle{-\sum_{\beta\neq 4}}\nu_{4\beta}\!\!\! & \nu_{45}\\
        
        \nu_{51} & \nu_{52} & \nu_{53} & \nu_{54} & \!\!\!\displaystyle{-\sum_{\beta\neq 5}}\nu_{5\beta}\!\!
\end{array} \right),
\end{equation}
The row summations vanish because $dT_\alpha/dt'=0$ whenever all of the components are in equilibrium at the same temperature.

The above-mentioned stiff solver is implemented in the code by alternating between the energy injection/photoionization step, and a thermalization step. The energy injection/photoionization step is a forward Euler step using Eqs.~(\ref{eqn:yHI}--\ref{eqn:Ej}), which form a system of $3N_{\mathrm{grid}}$ ordinary differential equations. The added energy is deposited into the electrons. The second step is a thermalization step, using the stiff integration method (Eq.~\ref{eqn:stiff}) to distribute energy among the species. To save computation time, ${\bf M}$ is computed using the temperatures at time $t'$.

To implement this, we need to compute the thermalization rates, i.e., the entries of $\mathbf{M}$. We discuss the interactions between different IGM components in the following four categories.

\subsection{Ionized + Ionized} \label{subsec:ionized-ionized}
Ionized particles include $e^-$, \HII, and \HeII. For two-body system, the corresponding energy transfer rate is  \citep{Anders1990} \begin{equation}
    \label{eqn:ionion}
    \nu_{\alpha\beta} = \frac{(m_\alpha m_\beta)^{1/2}Z_\alpha^2 Z_\beta^2 n_\beta \ln\Lambda_{\alpha\beta}}{(m_\alpha T_\beta + m_\beta T_\alpha)^{3/2}} \cdot 1.8\times10^{-19}\ \mathrm{s^{-1}},
\end{equation}
where $m_{\alpha/\beta}$ are the particle mass of $\alpha$ and $\beta$ respectively, $Z_{\alpha/\beta}$ are their atomic numbers, $n_\beta$ is the particle density of $\beta$, and temperature is in eV. We fix the value of the Coulomb logarithm $\ln\Lambda_{\alpha\beta} = \ln(b_{\mathrm{{max}}}/b_{\mathrm{{min}}})$ to a typical value, $\ln\Lambda \approx 28$, since it varies slowly with parameters. This value is taken from setting the maximum impact parameter $b_{\mathrm{{max}}}$ equal to the Debye length, and the minimum to $b_{\mathrm{{min}}} = e^2 / (2\pi\epsilon_0 m_e v^2_e)$ \citep{1983nrl..reptQ....B} at a temperature of $10^4\,$K.

\subsection{Ions + Neutrals} \label{subsec:ions-neutrals}
For ions (\HII\ and \HeII, denoted as $\alpha$) and neutrals (\HI\ and \HeI, denoted as $\beta$), their mutual interactions are primarily caused by the the polarization of neutrals induced by the electric field of ions, and resonant exchange for species with same atomic number. The energy transfer rate $\nu_{\alpha\beta}$ is related to the interaction cross section $\sigma_{\alpha\beta}$ through
\begin{equation}
    \label{eqn:ionsneutrals_rate}
    \nu_{\alpha\beta} = n_\mathrm{\beta} \frac{2m_\alpha}{m_\alpha + m_\beta} \nu_{\alpha\beta, p},
\end{equation}
where $n_\mathrm{\beta}$ is the neutral density, $\nu_{\alpha\beta,p} \equiv \langle \sigma_{\alpha\beta}v \rangle$ is the momentum transfer rate, $v$ is relative speed, and the factor of mass ratio is introduced by the conversion from momentum to energy transfer rate.

Assuming the polarization potential is in an ideal $r^{-4}$ form, the momentum transfer rate coefficient is a constant \citep{dra11}. For resonant exchange (\HI\ + \HII, \HeI \ + \HeII), we fit the total cross section as a function of temperature in a power law form, using data from \citet{1977RSPSA.353..575H} and \citet{2017PhPl...24f3502M}. 

\subsection{Electrons + Neutrals} \label{subsec:electrons-neutrals}
For interactions between electrons and neutrals, we also use a power-law fitting for the elastic collisions cross section. Experimental data is measured by \citet{1958PhRv..112.1157B} and \citet{1984PRA...30...1247G}.

\subsection{Neutrals + Neutrals} \label{subsec:neutrals-neutrals}
Due to the van der Waals interaction caused by dipole moment fluctuations, neutrals repulse each other at small range and weekly attract at larger separation. We do the power-law fitting for the elastic scattering cross section according to the measurements from  \citet{1973PhRvA...7...98G}.

\begin{figure}
\plotone{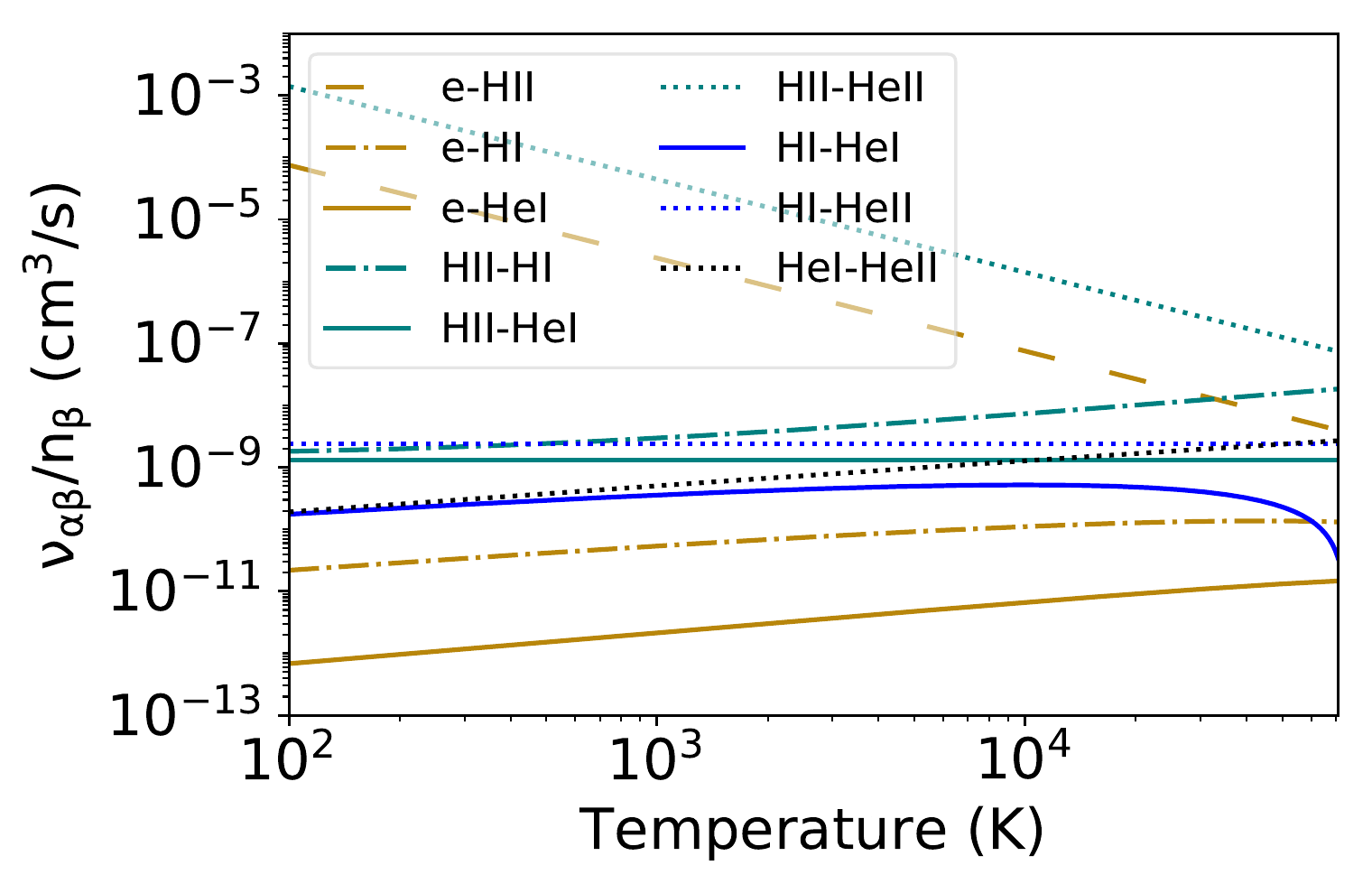}
\caption{The interaction rates between different IGM species, interpolated from the experimental data listed in sections \ref{sec:stiff}. We do not present e-\HeII\ interaction in the plot because it differs from e-\HII\ interaction rate only by a few percents and the two curves overlap in the figure.\label{fig:cross_section}}
\end{figure}

The interpolated interaction rates between IGM pairs as a function of temperature are shown in Figure \ref{fig:cross_section}. Electron-ion and ion-ion interactions dominate in low-temperature region (less than $10^4$ K) and decrease to become comparable with electron/ion-neutral rate.

\section{Results} \label{sec:result}

\begin{figure}
\plotone{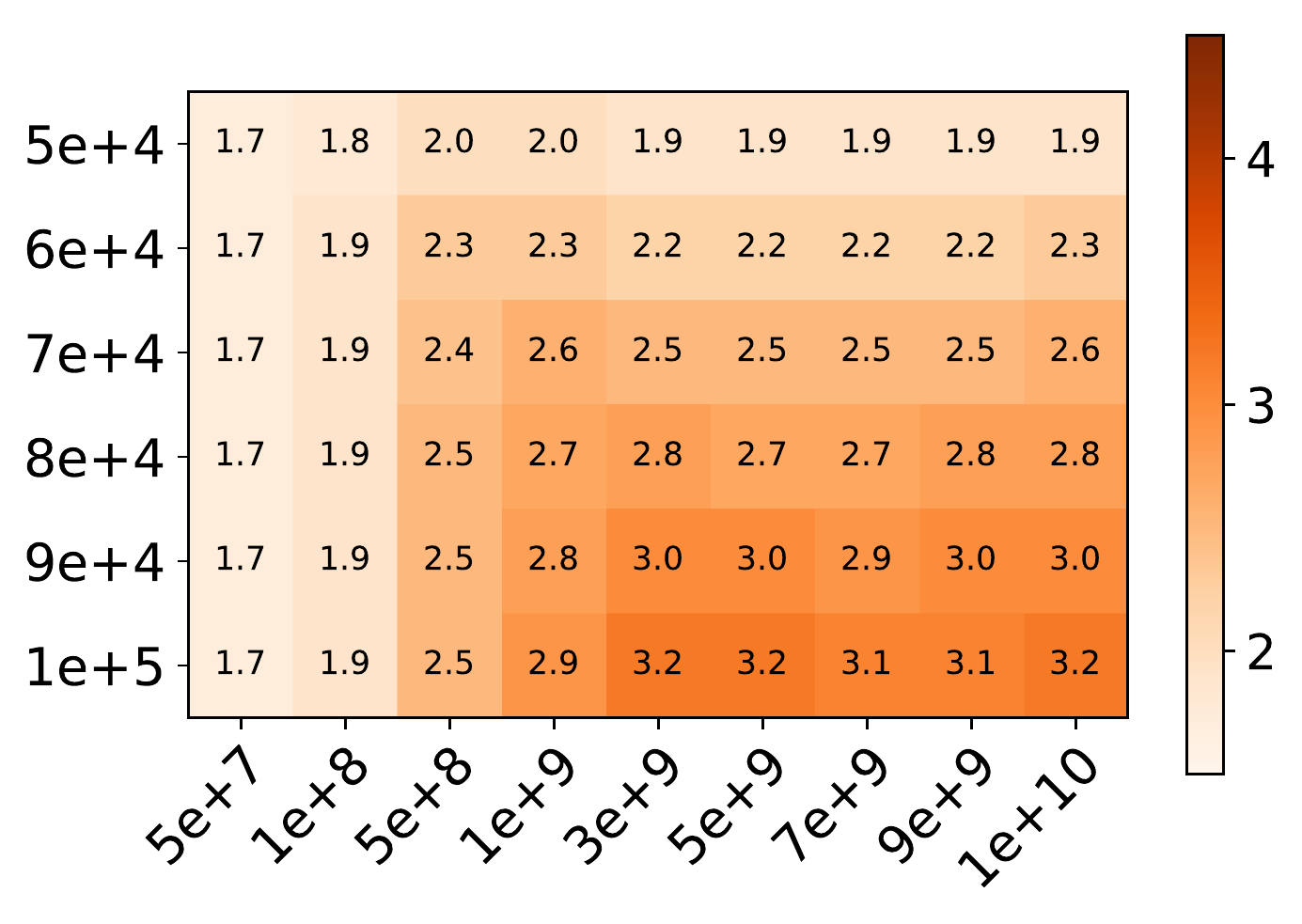}
\caption{$T_{\mathrm{re}}$ in $10^4\ $K as a function of ionization photon temperature $T_{\mathrm{bb}}$ (vertical, in K) and front velocity $v_i$ (horizontal, in $\mathrm{cm/s}$). Final temperatures shown here are for the non-equilibrium case.\label{fig:heatmap_Tre}}
\end{figure}

\begin{figure*}
\gridline{\fig{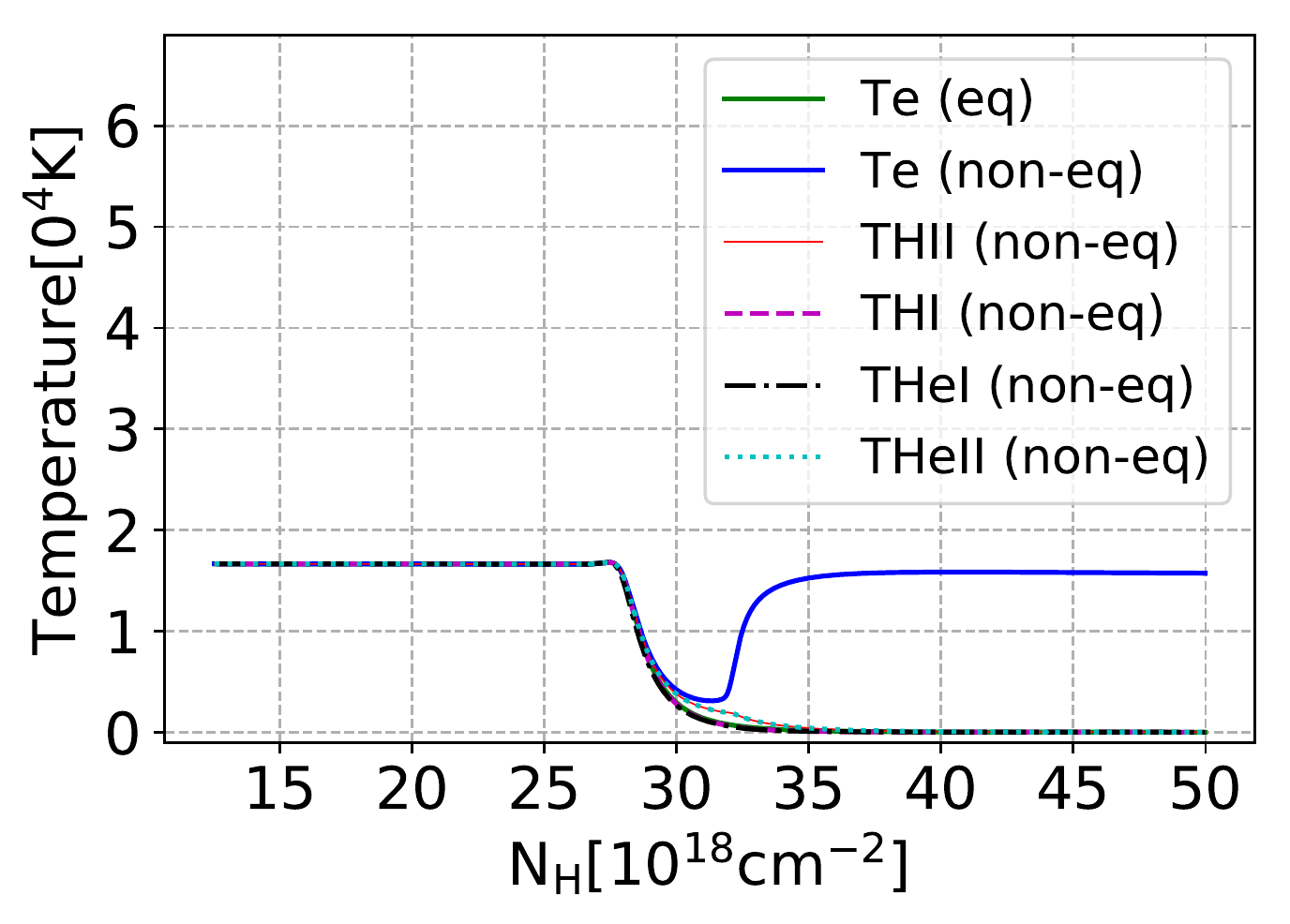}{0.333\textwidth}{(a) $T_{\mathrm{bb}} = 5\times10^4\ \mathrm{K}$, $v_i = 5\times10^7\ \mathrm{cm/s}$}
          \fig{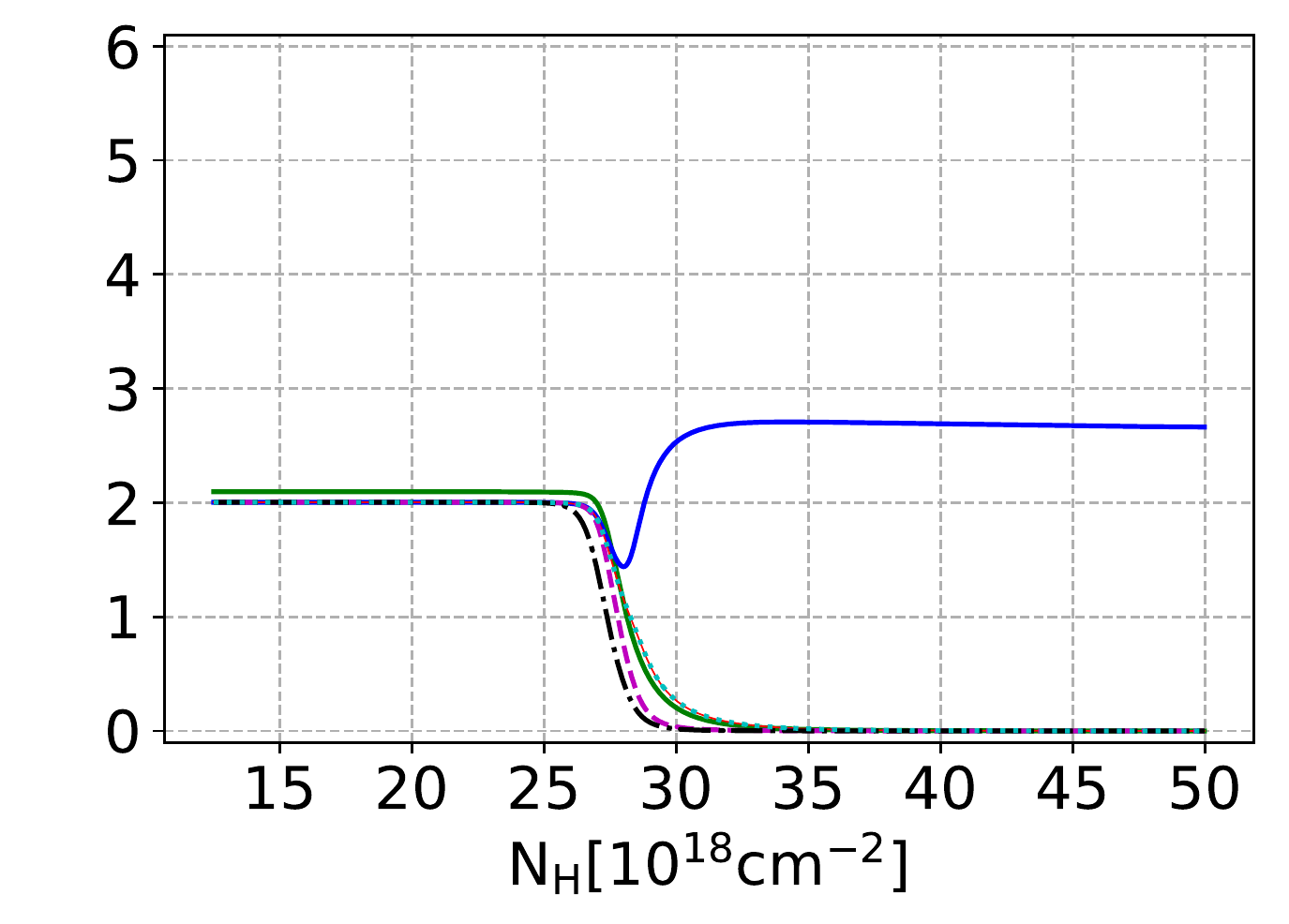}{0.333\textwidth}{(b) $T_{\mathrm{bb}} = 5\times10^4\ \mathrm{K}$, $v_i = 5\times10^8\ \mathrm{cm/s}$}
          \fig{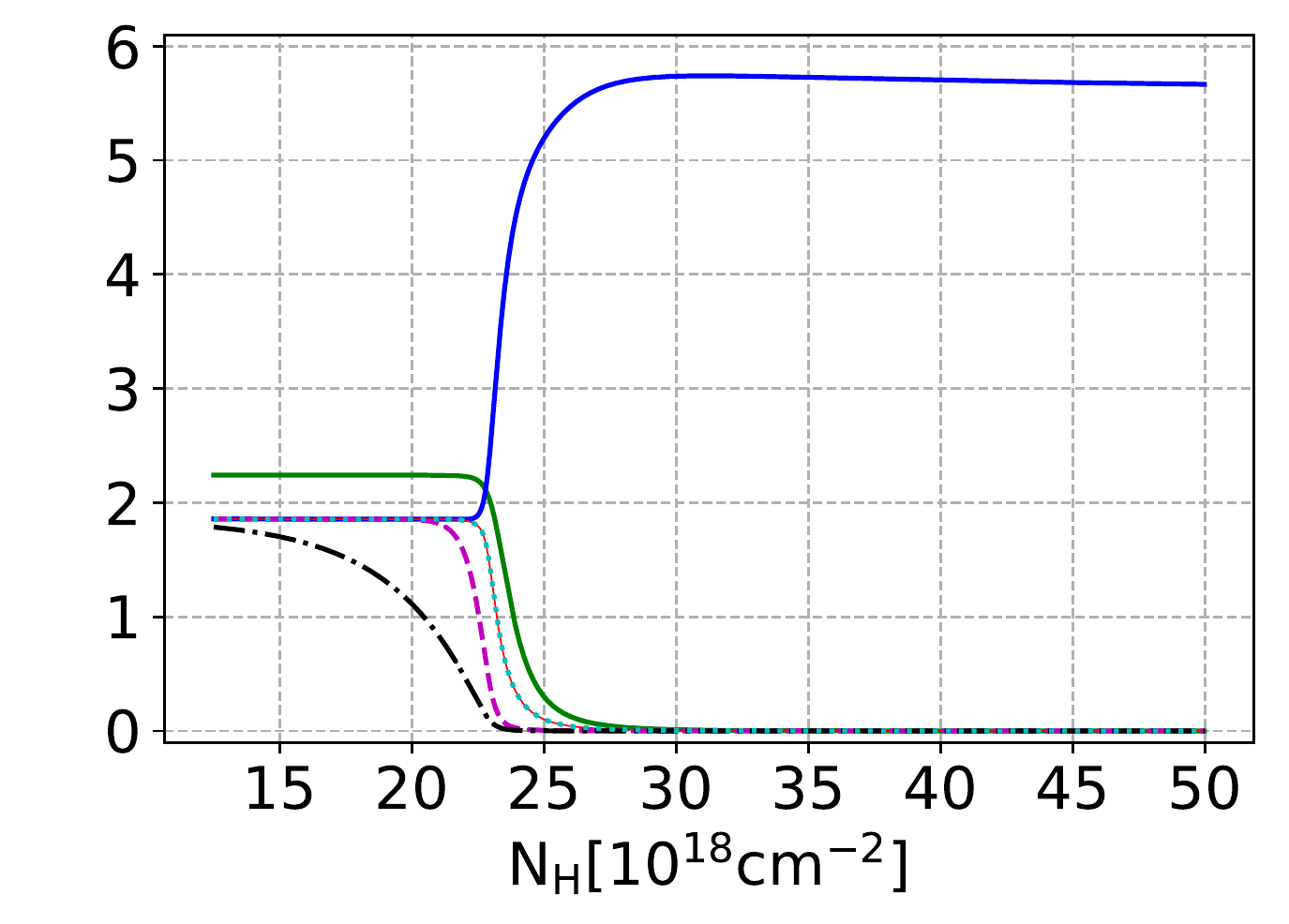}{0.333\textwidth}{(c) $T_{\mathrm{bb}} = 5\times10^4\ \mathrm{K}$, $v_i = 5\times10^9\ \mathrm{cm/s}$}
          }
\gridline{\fig{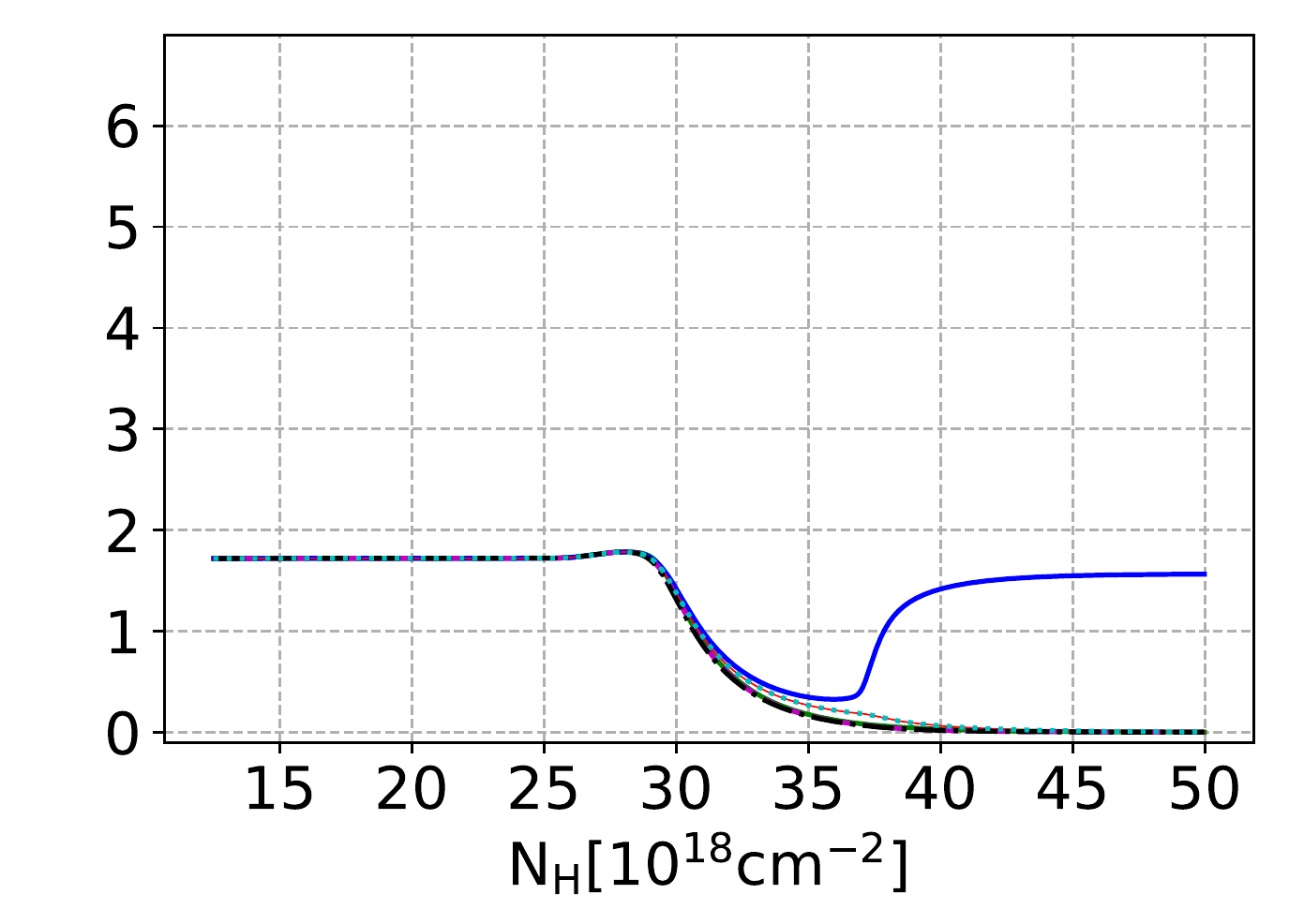}{0.333\textwidth}{(d) $T_{\mathrm{bb}} = 8\times10^4\ \mathrm{K}$, $v_i = 5\times10^7\ \mathrm{cm/s}$\label{fig:temp_evo_d}}
          \fig{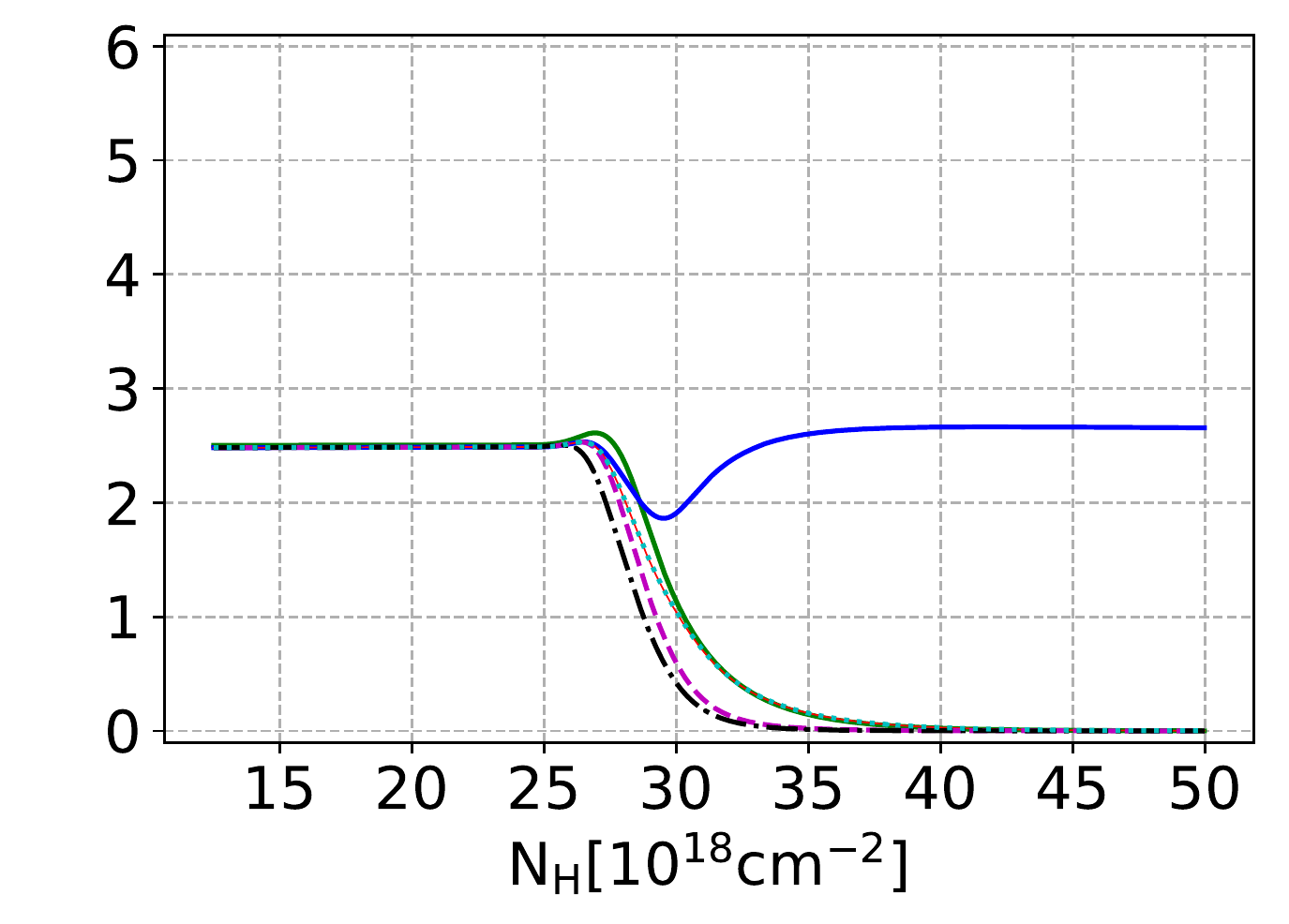}{0.333\textwidth}{(e) $T_{\mathrm{bb}} = 8\times10^4\ \mathrm{K}$, $v_i = 5\times10^8\ \mathrm{cm/s}$}
          \fig{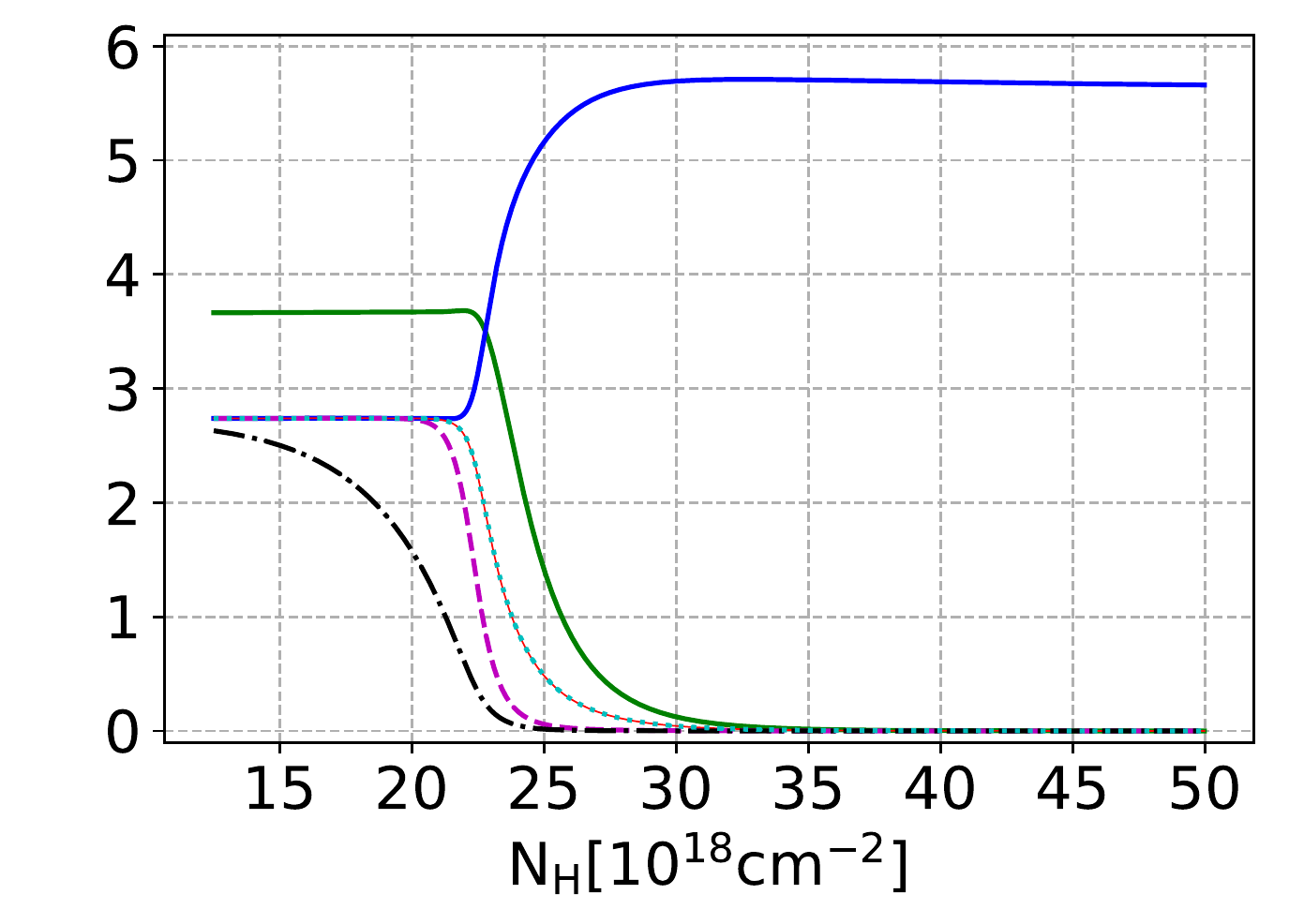}{0.333\textwidth}{(f) $T_{\mathrm{bb}} = 8\times10^4\ \mathrm{K}$, $v_i = 5\times10^9\ \mathrm{cm/s}$}
          }  
\caption{Temperature evolution of species at ionization front with different incident blackbody radiation fields $T_{\mathrm{bb}}$ and front velocities $v_i$. $T_{\mathrm{re}}$ refers to the plateau at $N_\mathrm{H} \lessapprox 25 \times 10^{18}\ \mathrm{cm}^{-2}$. Solid green curves assume IGM is always in equilibrium state, and other curves represent the temperature of each species when mutual interactions are included. In general, the non-equilibrium effects become stronger at higher velocity and temperature because the interaction rate gets smaller according to Equation \ref{eqn:equi}. For these typical incident radiations and front velocities, the final temperatures with interactions get lowered compared with the equilibrium case. \label{fig:temp_evo}}
\end{figure*}

\begin{figure}
\plotone{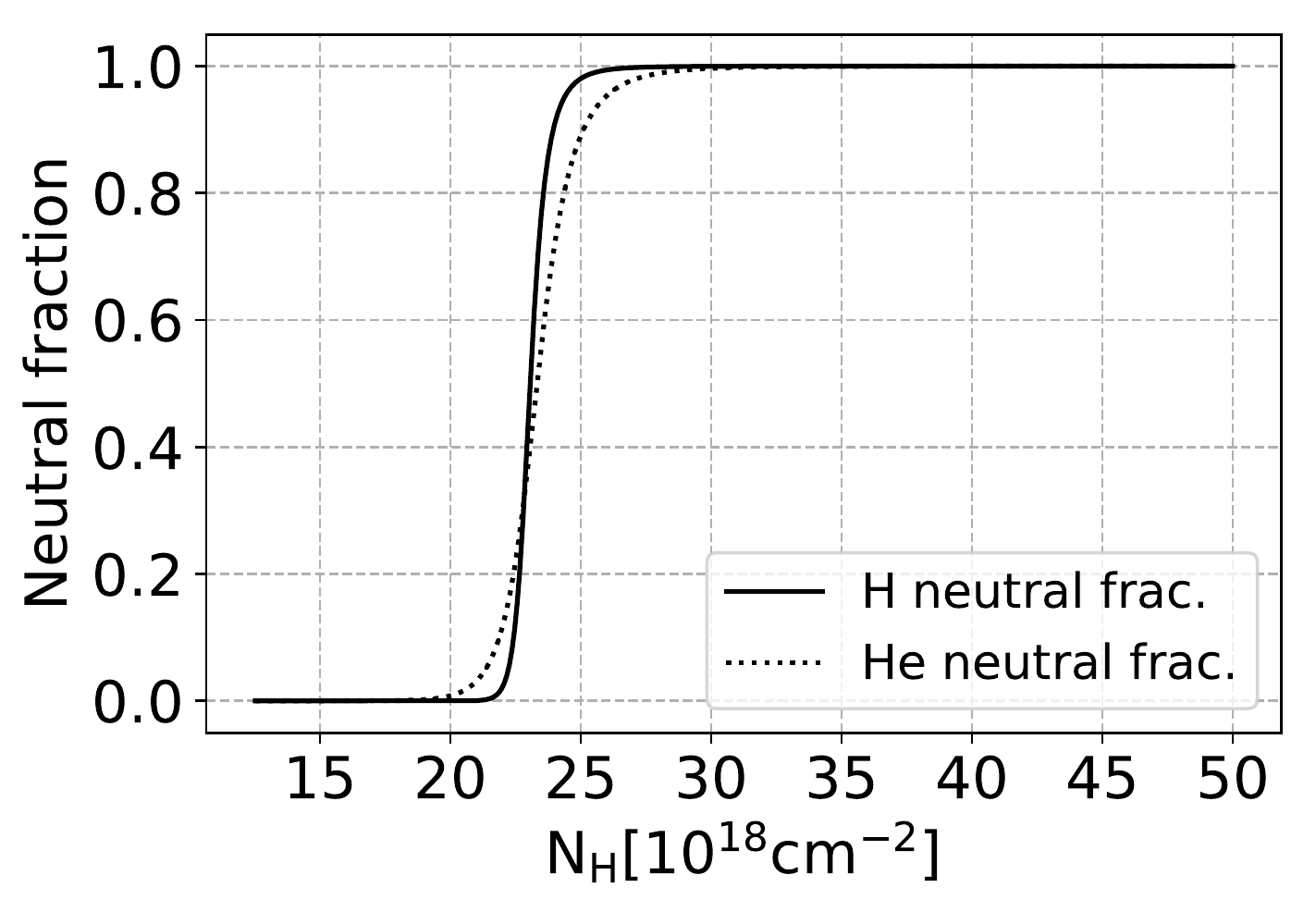}
\caption{Neutral fraction of hydrogen and helium with $T_{bb} = 5 \times 10^4$K and $v_i = 5 \times 10^9 \mathrm{cm/s}$.\label{fig:neutral_5}}.
\end{figure}

\begin{figure}
\plotone{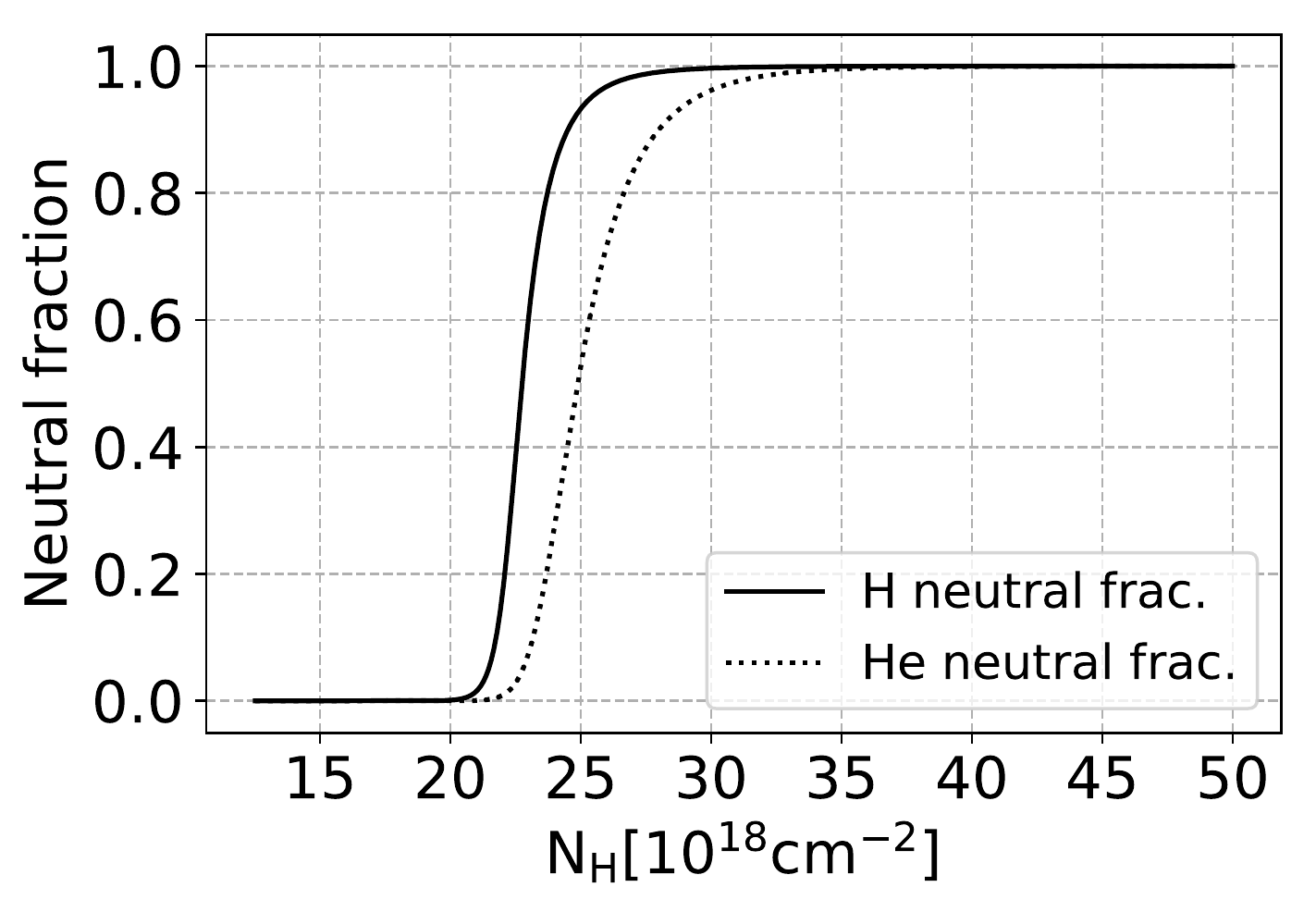}
\caption{Neutral fraction of hydrogen and helium with $T_{bb} = 8 \times 10^4$K and $v_i = 5 \times 10^9 \mathrm{cm/s}$ K.\label{fig:neutral_8}}.
\end{figure}

\begin{figure}
\plotone{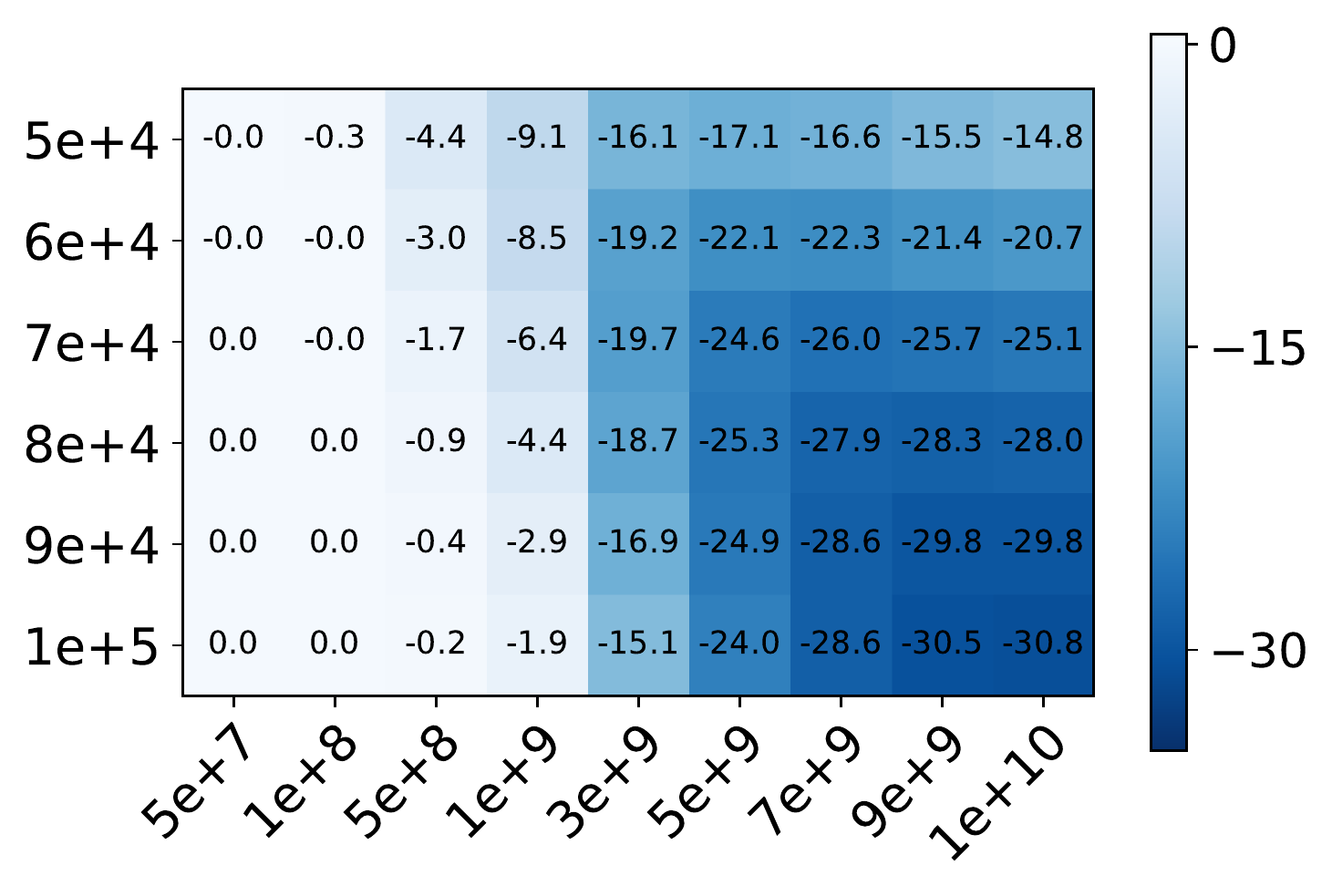}
\caption{Relative difference (in percentage) between $T_{\mathrm{re}}$ with and without non-equilibrium effects. Blue color indicates that interactions decrease $T_{\mathrm{re}}$. The vertical axis is the incident photon temperature in K assuming blackbody spectrum, and the horizontal axis represents I-front velocity in $\mathrm{cm/s}$. \label{fig:heatmap}}
\end{figure}

We aim to quantify the dependence of $T_{\mathrm{re}}$ on the incident spectrum temperature $T_{\mathrm{bb}}$ and front velocity $v_i$ when non-equilibrium interactions are present. The results are summarized from Figures \ref{fig:heatmap_Tre}--\ref{fig:heatmap}. In Figure \ref{fig:heatmap_Tre}, $T_{\mathrm{re}}$ gets higher with higher velocity (lower density) because of the smaller ratio of the Lyman-$\alpha$ cooling process to photoionization heating.

Figure~\ref{fig:temp_evo} illustrates a few examples of the equilibration process for a range of model parameters. The ionizing sources inject radiation from the left of the plot, and the ionization front propagates towards right. High-energy photons stream ahead of the ionization front, resulting in a temperature increase. The energy of photoionization goes into the electrons, with the result that the electrons in the mostly-neutral region (right side of the plot) are hotter than the ions or neutrals. Their temperature may be quite high ($> 5\times 10^4$ K in panel c) but because of the low ionization fraction they carry little thermal energy.

In general, all species stay in equilibrium at fully-ionized region (roughly $N_\mathrm{H} \lessapprox 25 \times 10^{18}\ \mathrm{cm}^{-2}$ in Figure \ref{fig:temp_evo}). Note that interaction rates are proportional to the density, so when written in terms of the re-scaled time they are inversely with $v_i$. Therefore, non-equilibrium effects are most important for the fastest ionization fronts. The sign of the effect is that in non-equilibrium, more of the thermal energy is in electrons that can cause Lyman-$\alpha$ cooling, thus compared with the equilibrium case (solid green curve in Figure \ref{fig:temp_evo}), $T_{\mathrm{re}}$ gets lowered when non-equilibrium interactions (the rest curves) are present. At the highest-velocity ($v_i \gtrapprox 10^8\ \mathrm{cm/s}$) region, temperatures of electrons and ions decouple from neutral species in the ionization front, and non-equilibrium effects become important. 


We can understand the non-equilibrium temperature of the electrons in the mostly neutral region by using order of magnitude arguments for the energy balance. As an example, consider the case of $T_{bb} = 8\times10^4$ K and $v_i = 5\times10^9\ \mathrm{cm/s}$ (Figure \ref{fig:temp_evo}f).

In the mostly neutral part of the front ($N_{\rm H}>3\times 10^{19}\,$cm$^{-2}$ in Figure \ref{fig:temp_evo}f), electrons reach a temperature of $\sim 5\times 10^4$ K, even though few ionizing photons have been absorbed. This is because the injection of energy into the electrons is fast and the electrons do not have time to transfer all of this energy to other species. The electron temperature in this region thus depends on the ionization front velocity $v_i$ and the average interaction rate $\bar{\nu}$ between electron and other species. Although there are incident photons between $1$ to $4$ Ry that contribute to the photoionization (at $>4$ Ry, the radiation is blocked by \ion{He}2), the medium to the right of the I-front is optically thick to low-energy photons, so that the photons that contribute to the heating are mostly the higher-energy photons ($\sim 4$ Ry). We estimate that the average photon energy deposit into an electron is $E_{e, \mathrm{inject}} = 2 .2$ Ry subtracting \HeI\ ionization energy from $4$ Ry. The energy present in the electrons is then related to the energy injected by these hard photons, and the ratio of the cooling time to the front time (the timescale on which this energy is injected):
\begin{equation}
	E_{e,\mathrm{remain}} = E_{e,\mathrm{inject}} \cdot \frac{\Delta t_{\mathrm{cool}}}{\Delta t_\mathrm{front}} = E_{e,\mathrm{inject}} \cdot \frac{\bar{\delta N}_{\mathrm{cool}}}{\bar{\delta N}_{\mathrm{front}}}.
\end{equation}
We take $\Delta t_{\mathrm{cool}} \equiv 1/(n_\mathrm{H} \bar{\nu})$ as the average time that electrons cool down due to collisional interactions with other species, and $\Delta t_{\mathrm{front}}$ as the e-folding time of the ionized electron fraction, which would be
\begin{eqnarray}
\bar{\delta N}_{\mathrm{front}} &\equiv& \frac1 {\sigma^{\mathrm{HI}} + f_\mathrm{He}\sigma^{\mathrm{HeI}}} = \frac1 {(0.12 + 0.079 \times 1.7) \times 10^{-18}} 
\nonumber \\
&\approx& 4.0 \times 10^{18}\ \mathrm{cm}^{-2}
\end{eqnarray}
using the \ion{H}1 and \ion{He}1 cross section at $4$ Ry. Then $\bar{\delta N}_{\mathrm{cool}} \equiv v_i \Delta t_{\mathrm{cool}}n_\mathrm{H} = v_i / \bar{\nu}$ is the average change of column density when electron transfer heats to the rest species, where $v_i = 5\times 10^9\ \mathrm{cm\ s^{-1}}$. The transfer rate coming from elastic collisions in a mostly neutral medium is $\bar\nu \sim \nu_{e,\rm HI}/n_{\rm H} 10^{-10}\,$cm$^3$/s (see Fig.~\ref{fig:cross_section}). However the actual cooling rate of the electrons is dominated by inelastic collisions where the hydrogen atoms are excited: for H($n=2$) cooling this is given by $\bar\nu \sim q_{1\rightarrow 2} E_{\rm Ly\alpha}/(\frac32kT_e)$, which is $(1.4,3.8,5.8)\times 10^{-9}$ cm$^3$/s at $T_e = (3,5,7.5)\times 10^4$\,K. This gives $\delta \bar N_{\rm cool} = (3.7,1.3,0.9)\times 10^{18}\,$cm$^{-2}$ and hence $2E_{e,\,\rm remain}/3k = (21,7.5,5.2)\times 10^4\,$K. Setting this equal to $T_e$, we see that there will be a solution between $T_e=5\times 10^4$ K and $7.5\times 10^4$ K; this is indeed consistent with the numerical result that $T_e = 6\times 10^4$ K in Fig.~\ref{fig:temp_evo}(f). We thus interpret the electron temperature in the mostly-neutral region to be the result of a balance between energy injection from photoionization by hard photons, and energy loss due to inelastic collisions that excite \HI\ and emit energy by Lyman-$\alpha$ and H(2s) two-photon emission.

In the temperature and velocity range of interests, ions (\HII \ and \HeII) stay in equilibrium all the time since their mutual interaction rate is high enough compared with the front propagation speed.

Naturally, the final temperature within the reionization bubble increases as the incident spectrum becomes harder ($T_{\mathrm{bb}}$ increases). However, the  importance of non-equilibrium effects exhibits a non-monotonic behavior with $T_{\mathrm{bb}}$ (Figure \ref{fig:heatmap}) at small I-front velocity ($v_i \lessapprox 5\times 10^9\ \mathrm{cm\ s^{-1}}$), when the collisional process overcomes radiative effects. This is because ion-neutral interaction and electron-ion interactions scale differently with temperature: ion/electron-neutral interaction rate scales with $T^{1/2}$ while electron-ion interaction rate scales with $T^{-3/2}$. Therefore, for low-temperature incident radiation ($T_{\mathrm{bb}} \lessapprox 7 \times 10^4\ \mathrm{K}$) electron-ion interaction rate is the larger in I-front, and electron/ion-neutral interaction becomes more significant at higher temperature ($T_{\mathrm{bb}} \gtrapprox 9 \times 10^4\ \mathrm{K}$). At $T_{\mathrm{bb}} \approx 8 \times 10^4\ \mathrm{K}$, both interaction rates are relatively low compared with timescale that particles stay in the I-front, so $T_{\mathrm{re}}$ decreases the most compared with the equilibrium case. At higher I-front velocity ($v_i \gtrapprox 5\times 10^9\ \mathrm{cm\ s^{-1}}$), Lyman-$\alpha$ and $\beta$ cooling, which are exponential in temperature, dominates over collisional process, so the final temperature decreases as $T_{\mathrm{bb}}$ goes up.

The bounds of the heat map are limited by our choice of energy sources for reionization. The upper limit of temperature we adopt ($10^5\ $K) corresponds to the effective blackbody temperature of the generic spectra for metal-free stars above $100 M_\sun$ \citep{2001ApJ...552..464B}. We have not considered even harder spectra, which might arise from ionization fronts dominated by AGN radiation. (See \citealt{2012MNRAS.426.1349M} and \citealt{2019ApJ...874..154D} for examples of such studies. Note that when X-ray radiation is included, one must also follow the energy loss of the secondary electrons prior to thermalization, as described in \citealt{2010MNRAS.404.1869F}.)

\section{Summary and Conclusions} \label{sec:conc}

We have presented a model of post-ionization-front temperatures $T_{\mathrm{re}}$, taking into account deviations from thermal equilibrium among the species in the IGM. Existing reionization simulations typically assume that photoelectrons thermalize baryon gas within an I-front in a timescale mush shorter than the time over which the baryon species stays inside the front.

To verify equilibration assumption and better predict $T_{\mathrm{re}}$, this paper made a first attempt to combine density-dependent ionization front model with an implicit stiff solver to include baryons elastic interactions. Our main results are as follows:
\newcounter{Xlist}
\begin{list}{\arabic{Xlist}.}{\usecounter{Xlist}}
\item
The equilibrium assumption is valid for ionization fronts slower than $10^9\ \mathrm{cm/s}$. However, baryon species start to thermally decouple at higher-velocity (lower-density) region, where the  momentum and energy transfer rates are comparable with the front speed. This can be illustrated by Figure \ref{fig:temp_evo} (c) and (f). Photoelectrons first heat up ions and \HI\ through elastic collisions, and \HeI\ is the last to be thermalized due to its small momentum transfer cross section.
\item
Adding non-equilibrium effects, this density-dependent model still predicts final temperature  $T_{\mathrm{re}}$ decreases as local density $\Delta$ decreases (or I-front velocity $v_i$ increases), because of the greater importance of the collisional cooling effects (Figure \ref{fig:heatmap_Tre}). $T_{\mathrm{re}}$ ranges from $1.7\times 10^4\ $K to $3.2\times 10^4\ $K in our parameter space.
\item
We demonstrated that during the EoR, the non-equilibrium interactions will affect $T_{\mathrm{re}}$ up to a level of $30\%$. Higher-velocity (lower-density) region has smaller momentum transfer rate, so baryons species have lower temperatures than photoelectrons in the ionization front. Therefore, photoelectrons will be averaged to a lower final temperature than the equilibrium case. Compared with the equilibrium model, $T_{\mathrm{re}}$ is decreased around twenty percents at the region $v_i \gtrapprox 10^9\ \mathrm{cm/s}$ (Figure \ref{fig:heatmap}), where $T_{\mathrm{re}}$ is approximately between $1.9\times 10^4\ $K and $3.2\times 10^4\ $K. $T_{\mathrm{re}}$ is lowered more rapidly as $v_i$ gets closer to the speed of light.
\end{list}

Future improvements include a better interpolation of the momentum transfer cross section and using a realistic incident spectrum (e.g. quasar spectrum) instead of blackbody, and a two-dimensional grid model to better describe inhomogeneous ionized bubble.

We have seen that non-equilibrium effects can substantially affect the structure of an ionization front, especially for the electron temperature. But from an astrophysical perspective, our main interest is in the implications for the temperature evolution of the IGM some time {\em after} reionization. While a full analysis of the temperature of the IGM, including its spatial variations, requires a hydrodynamic simulation, we can assess the likely impact using the post-reionization temperature evolution model of \citet{2016MNRAS.460.1885U}. If we consider a region that was reionized at mean density at $z_{\rm re} = 8$ by an ionization front at $v_{\rm i} = 5\times 10^8$ cm/s and a spectrum with $T_{\rm bb}=5\times 10^4$ K, our model predicts a reionization temperature of $T_{\rm re} = 2.0\times 10^4$ K and a 4.4\% difference between the equilibrium and non-equilibrium temperatures at $z_{\rm re}$. This difference falls to 1.8\% by $z=5.5$ and 1.6\% at $z=4.0$ due to the convergent nature of temperature evolution. If we have an ionization front velocity ten times larger, $5\times 10^9$ cm/s, and consider a parcel of gas reionized at $z_{\rm re} = 7$, then the temperature difference between equilibrium and non-equilibrium calculations is 17\% at $z=z_{\rm re}$, and is still 9\% at $z=5.5$ and 7\% at $z=4.0$. These corrections are smaller than, but not entirely negligible compared to, the observational errors in IGM temperature determination at $z\sim 4$ \citep[e.g.][]{2011MNRAS.410.1096B}.

\section{Acknowledgement} \label{sec:ack}
We thank Anson D'Aloisio, Xuelei Chen, Kevin Ingles, Ely Kovetz, Paulo Montero-Camacho, and Yan Gong for very useful discussions.

CMH is supported by the Simons Foundation award 60052667 and NASA award 15-WFIRST15-0008.

\end{document}